\documentclass[8pt,floatfix,twocolumn,showpacs,
preprintnumbers,amsmath,amssymb,amsfonts,superscriptaddress,prb]{revtex4}

\usepackage{graphicx,bm}
\usepackage{dcolumn}

\newcommand{\zSc}{$\langle {\bm\tau}_1\cdot\mathbf S_1\rangle$}
\newcommand{\SSc}{$\langle \mathbf S_1\cdot\mathbf S_5\rangle$}
\newcommand{\zScO}{$\langle {\bm\tau}_3\cdot\mathbf S_3\rangle$}
\newcommand{\SScO}{$\langle \mathbf S_3\cdot\mathbf S_6\rangle$}

\newcommand{\JKt}{$J_{\rm K}/t$}
\newcommand{\JK}{$J_{\rm K}$}

\begin{document}

\title{Kondo Lattice Model with Finite Temperature Lanczos Method}
\author{I. Zerec}
\affiliation{Max-Planck-Institute for the Physics of Complex Systems,
D-01187 Dresden, Germany}
\author{B. Schmidt}
\affiliation{Max-Planck-Institute for Chemical Physics of Solids,
D-01187 Dresden, Germany}
\author{P. Thalmeier}
\affiliation{Max-Planck-Institute for Chemical Physics of Solids,
D-01187 Dresden, Germany}
\date{\today}

\begin{abstract}
We investigate the Kondo Lattice Model on 2D clusters
using the Finite Temperature Lanczos Method. The temperature dependence of thermodynamic and correlations functions are systematically studied for various Kondo couplings \JK. The ground state value of the total local moment is presented as well. Finally, the phase diagrams of the finite clusters are constructed for periodic and open boundary conditions. For the two boundary conditions, two different regimes are found for small \JKt, depending on the distribution of non-interacting conduction electron states. If there
are states within $J\rm_K$ around the Fermi level, two energy scales,
linear and quadratic in \JK, exist. The former is
associated with the onsite screening and the latter with the RKKY interaction.
If there are no states within \JK{} around the Fermi level, the only
energy scale is that of the RKKY interaction. Our results imply that
the form of the electron density of states (DOS) plays an
important role in the competition between the Kondo screening and the RKKY
interaction. The former is stronger if the DOS is larger around the Fermi
level, while the latter is less sensitive to the form of the DOS.
\end{abstract}


\maketitle

\section{Introduction}

The Kondo lattice model (KLM) is used to describe compounds containing
localized magnetic moments, such as heavy fermion systems and Kondo
insulators.~\cite{Tsunetsugu:1997} The KLM Hamiltonian is given by:
\begin{equation}\label{KLH}%
H_{\rm KL}= - t\sum_{i,j;\sigma}%
c_{i,\sigma}^\dagger c_{j,\sigma} +%
J_{\rm K}\sum_{i}{\mathbf {\bm\tau}_{i}}\cdot {\rm \bf S}_i.%
\end{equation}%
Here, ${\bm\tau}_i$ and ${\bf S}_i$ are itinerant and local ($f$-)
spins on site $i$, respectively. The KLM takes into account hopping of 
conduction electrons, $t$, and their Kondo interaction, $J_{\rm K}$,
with local $f$-spins. 
The Kondo term causes a screening of local spins, but
also induces indirect RKKY interactions between the local spins on the
lattice. These two interactions compete, leading to either magnetically
ordered or non-magnetic ground states, separated by the quantum
critical point (QCP).~\cite{Doniach:1977} The nature of the QCP in
heavy fermion compounds is one of the central topics of condensed matter
physics today.~\cite{Rech:2005,Sun:2005,Paschen:2004,Custers:2003,Si:2001,Schroder:2000,Mathur:1998,Steglich:1996}

Various analytical and numerical methods
have been used to study the KLM, like the mean-field approach~\cite{Zhang:2000a} and
Quantum Monte Carlo (QMC) method.~\cite{Capponi:2001} The finite temperature Lanczos method (FTLM)~\cite{Jaklic:2000} has been used to study thermodynamic 
functions.~\cite{Haule:2000} Also the periodic Anderson
model on finite size clusters has been studied with similar methods
recently.~\cite{Luo:2005}

Here, we use the FTLM and mainly focus
the analysis on the temperature dependence and the ground state values of
static correlation functions, of central importance for the QCP.  We compare onsite vs. intersite correlations on the lattice and also compare the former with the impurity case. The latter is associated with RKKY interactions. We also investigate the evolution of the total onsite moment with Kondo coupling strength \JK. The main advantage of the FTLM is that 
it involves no uncontrolled approximations and is exact for finite
size clusters. It treats both, Kondo screening and RKKY
interactions, correctly. 

\begin{figure}
\begin{center}
\includegraphics[width=0.99\linewidth,clip=]{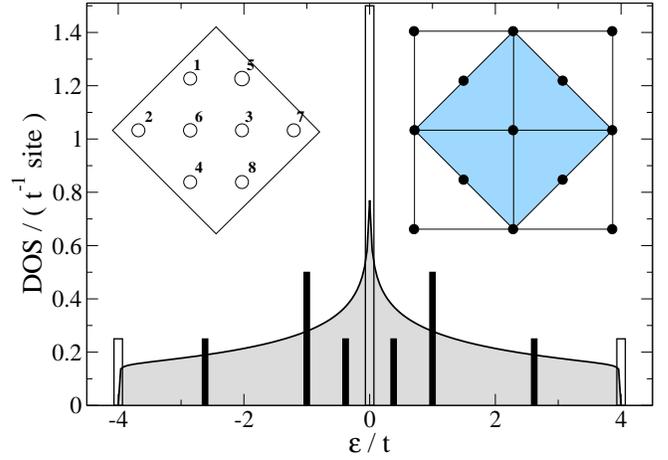}
\caption{\label{dos} DOS for the tight-binding model in 2D is
shown shaded. Unshaded bars represent discrete levels of the 2D
8-site cluster (for $J\rm_K=0$) with PBC. The black bars represent
levels with OBC. The bar heights correspond to the level degeneracy 
(divided by 8). Left Inset: The 8-site cluster with site 
indices. Right Inset: The first Brillouin zone with the k-points
corresponding to Bloch states of 8-site cluster. The shaded part
represents the Fermi volume for half filled case. There are 6 degenerate
states on the "Fermi surface" when the multiplicities are taken into
account correctly.}
\end{center}
\end{figure}

We consider the 8-site cluster of a square lattice, as shown in the inset of
Fig.~\ref{dos}, with periodic (PBC) and open (OBC) boundary conditions
at half filling.

We will show that for this cluster the two different boundary
conditions lead to two interesting physical situations, which may have
relevance to the two types of QCP's observed (cf. Ref.~\onlinecite{Si:2001} and references cited therein). There is of course only one solution of the KLM in the thermodynamic limit. However, the real materials have more complex band structures than described by the n.~n. hopping in~(\ref{KLH}). By varying boundary
conditions for the KLM on the small cluster one may mimic
band structures with different characteristic behaviour of the
density of states (DOS) around the Fermi level. For this reason we present in
detail the analysis of the KLM on the finite cluster for the PBC and
the OBC and compare the two cases. In Sec. II we discuss the specific
heat anomalies obtained in the two cases. Sec. III discusses the
central problem of Kondo-lattice physics: The competition and
crossover behaviour of onsite and intersite correlations as function
of temperature and control parameter \JKt. The associated Kondo
screening behaviour evident from the susceptibility and total local
moment is presented in Sec. IV. Finally Sec. V summarizes the results
in a phase diagram and gives the conclusions.

\section{Specific heat}

\begin{figure}
\begin{center}
\includegraphics[width=0.9\linewidth,clip=]{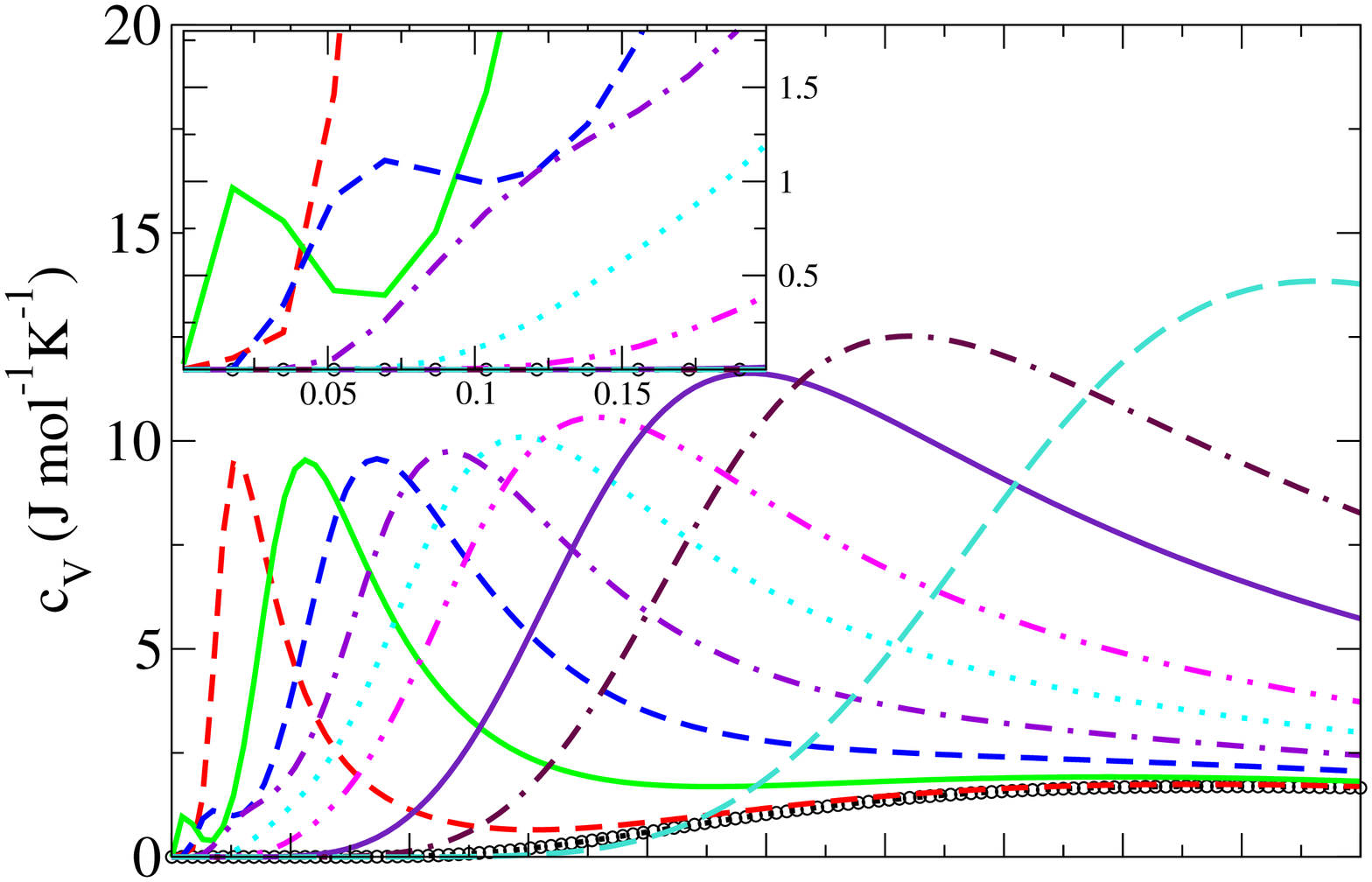}
\includegraphics[width=0.9\linewidth,clip=]{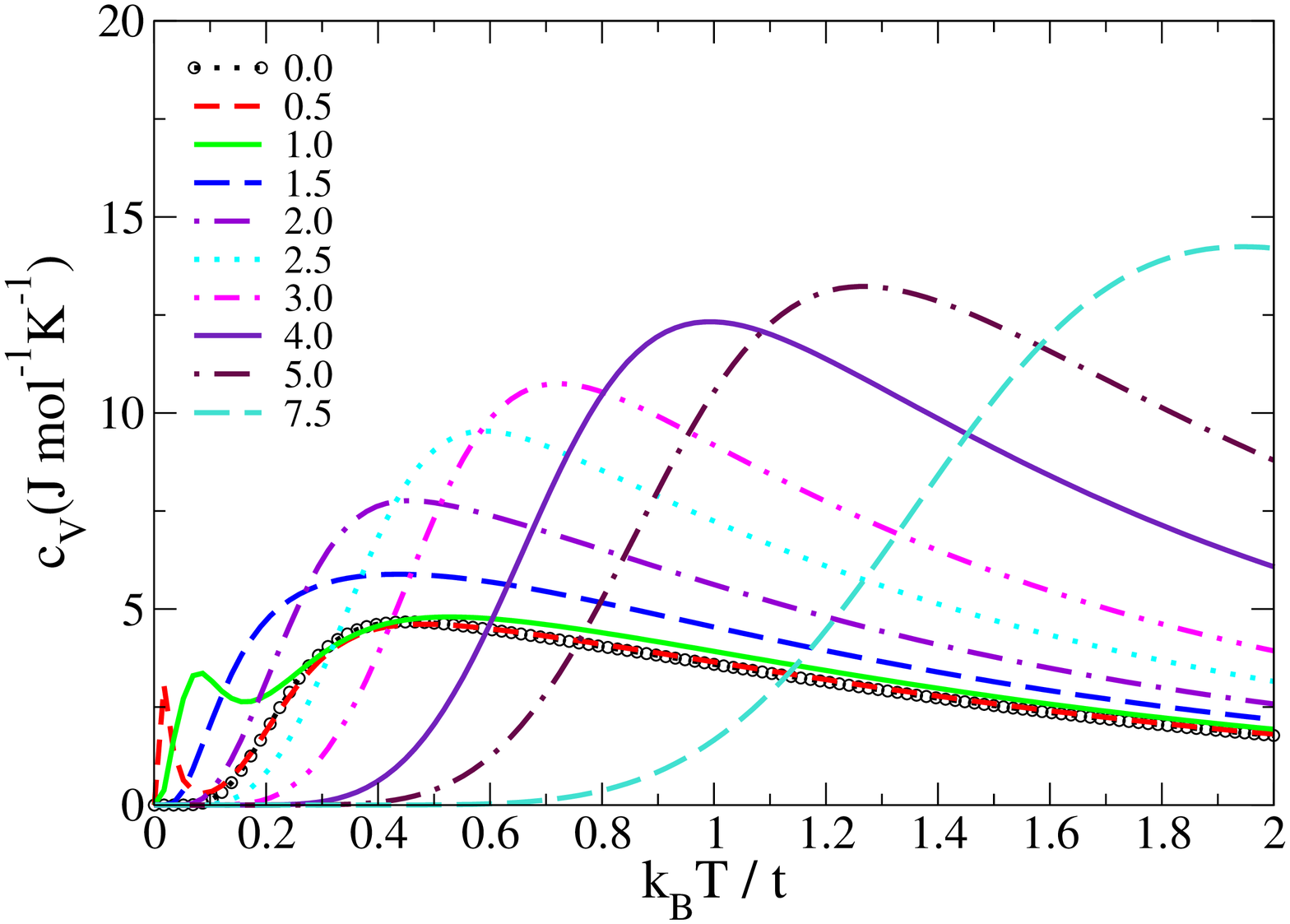}
\caption{\label{cV} (Color online) The specific heat calculated for
various values of \JKt{}, as indicated in the legend. The legend is valid for all
the figures in this work, if not explicitly specified. Above: for PBC. The two low temperature peaks can be seen for $0.5<$\JKt$\lesssim2$. In the inset the lower peaks are magnified. Below: for OBC. The \JKt$=0$ curve agrees well with the thermodynamic limit of the Hamiltonian~(\ref{KLH}). For \JKt$>0$ there is only one low temperature peak.}
\end{center}
\end{figure}

The specific heat for the PBC and the OBC, for various coupling constants \JK{},  is shown in Fig.~\ref{cV}. 

The low-temperature specific heat for non-interacting case,
$J_{\rm K}=0$, with PBC deviates significantly from the thermodynamic
limit. The reason is that 6 out of 8 electrons are placed on the "Fermi
surface", shown in the inset of Fig.~\ref{dos}, thus the ground state is
highly degenerate. The
specific heat for OBC is closer to the 
thermodynamic limit. The peak from the hopping term for $J\rm_K=0$ is 
much more pronounced and at lower temperature, $T_{\rm max,0}^{\rm OBC}$,
for OBC than for PBC.
The electronic states for OBC are no longer Bloch
states, but the important point is that the electron energies are shifted
from, and distributed below, the "Fermi level". The distribution of these
discrete levels is such that it relatively well mimics the continuum
DOS in the thermodynamic limit, as shown in Fig.~\ref{dos}.

The specific heat for PBC and OBC, when Kondo interaction is turned on,
shows qualitatively different behavior in the weak coupling regime,
$J_{\rm K}/t\lesssim2$, as shown in Fig.~\ref{cV}. For PBC, a double peak
structure 
is observed at low temperatures for $0.5<J_{\rm K}/t\lesssim2$. We may denote the positions of the
upper and lower peak by $T\rm_{max,1}^{PBC}$ and $T\rm_{max,2}^{PBC}$,
respectively. For OBC only one low temperature 
peak appears at $T\rm_{max,1}^{OBC}$. 
The origin of these peaks are spin degrees of freedom,
whose degeneracy is lifted by Kondo interaction. In the strong coupling limit there is only one peak in the specific heat, and it does not depend on the boundary conditions.

The specific heat for OBC agrees with the results obtained with the same method for the 10-site cluster with PBC in Ref.~\onlinecite{Haule:2000}, as well as with the results obtained with the QMC in Ref.~\onlinecite{Capponi:2001}.

\begin{figure}
\begin{center}
\includegraphics[width=0.9\linewidth,clip=]{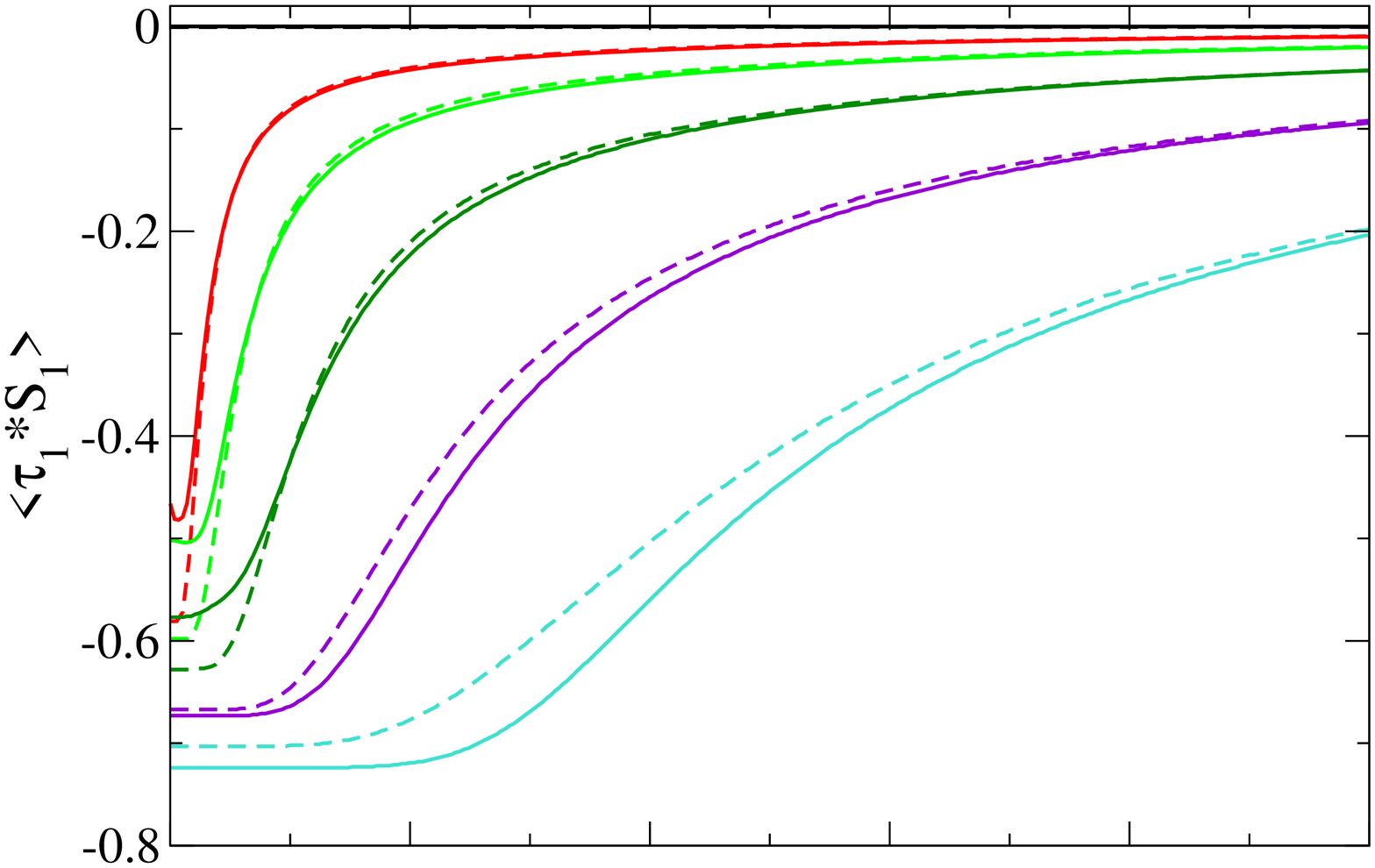}
\includegraphics[width=0.9\linewidth,clip=]{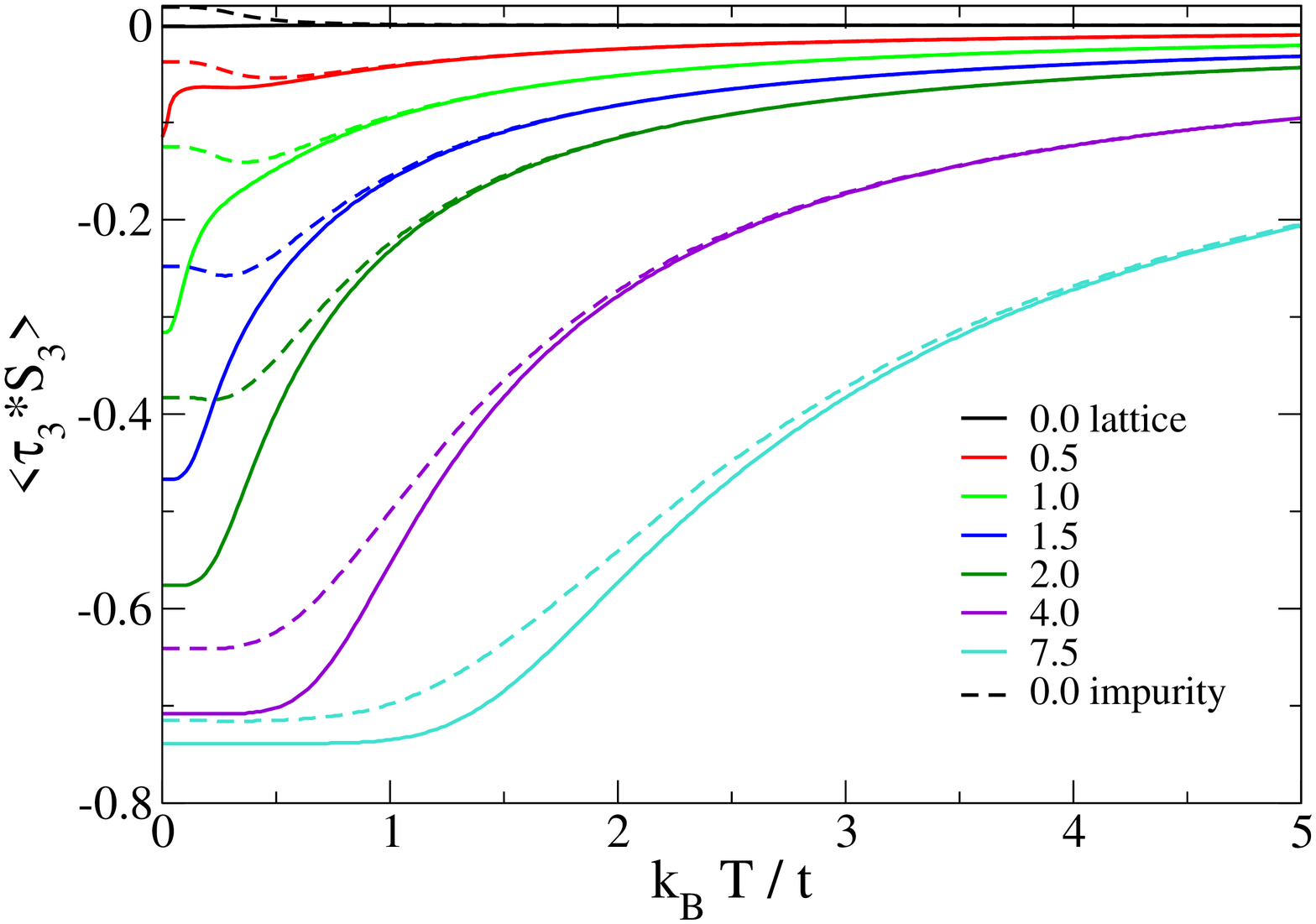}
\caption{\label{corrT} (Color online) The onsite correlations in the
lattice (full lines) as function of temperature and the comparison
with the impurity case (broken lines), for \JKt{} values as indicated in the legend. The correlation strength increases monotonically with increasing \JKt. In the impurity case, the impurity local spin is placed on site 1 for the PBC, and site 3 for the OBC. Above: for PBC. Below: for OBC. Due to the small system size for the impurity case, there are some unphysical deviations at low temperatures and weak coupling \JKt{} for the OBC.}
\end{center}
\end{figure}

\section{Spin correlation functions: onsite singlet formation
vs. RKKY interactions}

\begin{figure}
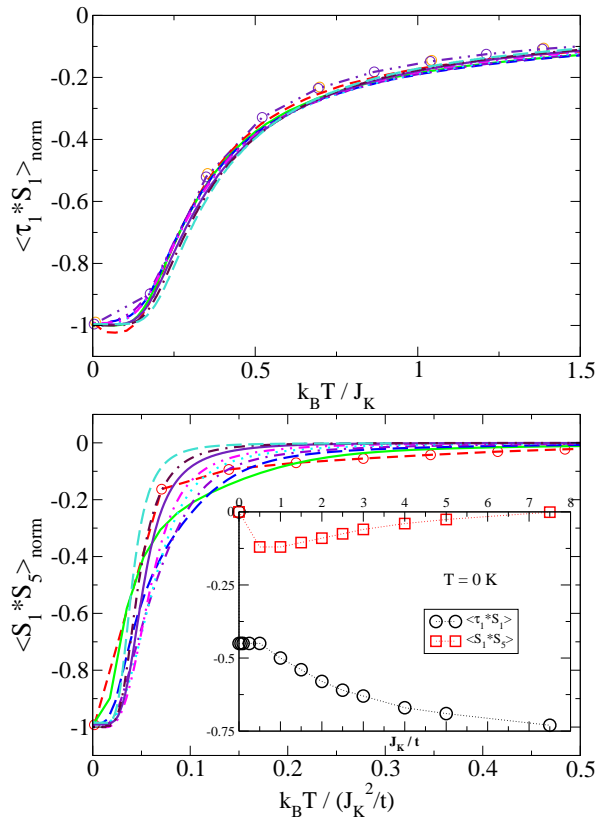

\begin{center}
\includegraphics[width=0.9\linewidth,clip=]{z1S1_sca_PBC_art.eps}
\includegraphics[width=0.9\linewidth,clip=]{S1S5_sca_PBC_art.eps}
\caption{\label{corrPBC} (Color online)  Correlation functions for PBC. Above:
Normalized \zSc{} as function of temperature divided by \JK. Below:
Normalized \SSc{} as function of temperature divided by 
the square of \JK. The normalization is achieved by dividing the curves with the ground state values. In the inset the ground state ($T=0$~K) correlations as function of \JKt{} are shown.}
\end{center}
\end{figure}

\begin{figure}
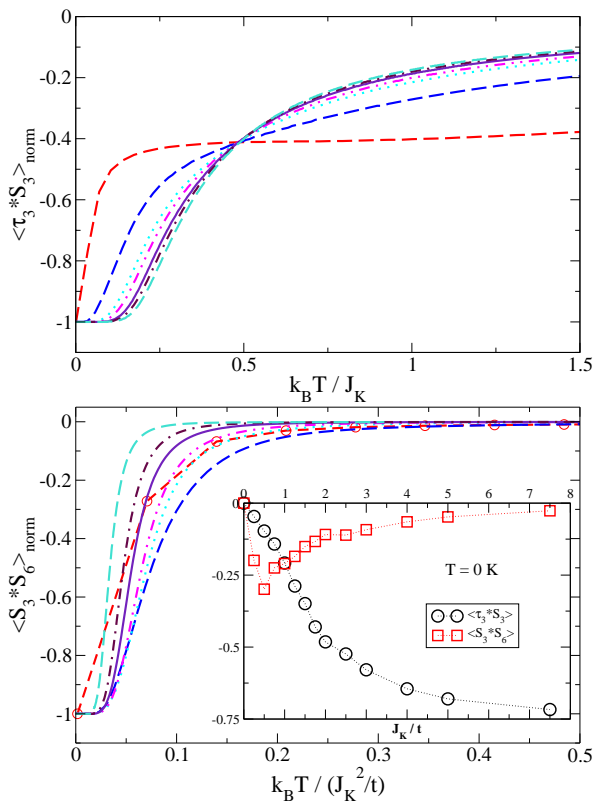

\begin{center}
\includegraphics[width=0.9\linewidth,clip=]{z3S3_sca_OBC_art.eps}
\includegraphics[width=0.9\linewidth,clip=]{S3S6_sca_OBC_art.eps}
\caption{\label{corrOBC} (Color online) Correlation functions for OBC. Above:
Normalized \zScO{} as function of temperature divided by \JK. Below:
Normalized \SScO{} as function of temperature divided by \JK$^2/t$. The normalization is achieved by dividing the curves with the ground state values. In the inset the ground state ($T=0$~K) correlations as function of \JKt{} are shown.}
\end{center}
\end{figure}

The nature of the specific heat peaks may be revealed by considering
appropriate static correlation functions. We focus the analysis on the following two:
\zSc{} is the onsite correlation function between local
and itinerant spins, while \SSc{} is the intersite correlation
function between local spins on neighboring sites for PBC. For OBC these are \zScO{} and \SScO{}, i.e., the inner sites 3 and 6 are considered, because they are more representative than sites on the edges of the cluster (cf. Fig.~\ref{dos}).

\subsection{Periodic Boundary Conditions}

The onsite correlations, \zSc, for PBC as function of temperature are shown in Fig.~\ref{corrT}. They are also compared with the onsite correlations for the impurity case, where there is only one local spin, at site 1, on the 8-site cluster. In the weak coupling regime the onsite screening is decreased in the lattice, because there are less itinerant spins per local spin available for the screening.

The inflection points of the correlation functions, as function of temperature, indicate that the upper
peak in the specific heat, at $T\rm_{max,1}^{PBC}$, corresponds to the formation of the onsite correlations,
whereas the lower one, at $T\rm_{max,2}^{PBC}$, 
corresponds to the formation of 
the intersite correlations. The ground state ($T=0$~K)
correlations for PBC as function of coupling strength \JKt{} are shown in the inset of Fig.~\ref{corrPBC}.
It is seen that the onsite correlations are stronger than the intersite correlations for all \JKt. Another characteristic is that \zSc{} jumps
to the large absolute value as soon as \JK{} is turned on. This is a consequence
of the degenerate itinerant degrees of freedom at Fermi level for
$J\rm_K=0$. The free itinerant spins on the "Fermi level" screen the local spins onsite as soon as $J\rm_K>0$; the ground state splits off and is
characterized by finite \zSc{} correlations. The unscreened
parts of the local spins form intersite correlations, \SSc{}, below
$T\rm_{max,2}^{PBC}$, due to the RKKY interaction. The magnitude of \SSc{} increases
rapidly as \JK{} is turned on. Because of the very small energy scale
(quadratic in \JK), convergence problems appear in the calculation of ground state \SSc{} correlations for $J_{\rm K}/t<0.5$ and we were not able to determine whether they also jump for finite \JK. In the strong coupling regime the intersite correlations vanish, while the onsite correlations saturate at $-3/4$, as the onsite singlets are formed.

The correlation functions, normalized to the ground state values, versus scaled temperature, as done in Ref.~\onlinecite{Capponi:2001}, are shown in
Fig.~\ref{corrPBC}. The onsite correlations are characterized by a
single temperature scale, perfectly linear in \JKt. The intersite
correlations are also characterized by a single temperature scale
proportional to \JK$^2/t$. This shows that for PBC the onsite
screening is independent of the intersite correlations, which may be
associated with RKKY interactions. This is a consequence of the
special symmetry of the 8-site cluster which implies a large fraction
of conduction electron sitting on the "Fermi surface". Nevertheless it
is an interesting result which may have relevance for a case where the
conduction electron DOS is strongly peaked at the Fermi level.

\subsection{Open Boundary Conditions}

The onsite correlations, \zScO, for OBC as function of temperature and the comparison to the impurity case are also shown in Fig.~\ref{corrT}. In contrast to the PBC, for OBC the onsite correlations are enhanced in the lattice. In the case of OBC there are no "free" itinerant spins, i.e., conduction electrons on the "Fermi level". In the weak coupling limit the onsite screening is not a direct process and is weak. The interesting point is the appearance of a new low energy scale in the lattice, which may be noted as a low-temperature downturn in the onsite correlations in Fig.~\ref{corrT}.

The ground state correlations are shown in the inset of
Fig.~\ref{corrOBC}.  The onsite correlations, \zScO{}, do not jump as
\JK{} is turned on, but rather increase continuously with \JK. The 
intersite correlations, \SScO{}, are much stronger than \SSc{} for PBC and in the weak coupling limit they are even larger in absolute values than the onsite \zScO{} correlations. It is understandable, because the local spins are only weakly screened, due to the lack of "free" itinerant spins. Therefore, the unscreened local spins can form strong intersite correlations.

The normalized correlations as function of the scaled temperature for
OBC are shown in Fig.~\ref{corrOBC}. The \zScO{} scale with \JK{}
only in the strong coupling limit. In the weak coupling limit two
characteristic temperatures or energy scales can be identified, as
already mentioned above. The upper one is associated with the energy
splitting of the discrete energy levels (cf. Fig.~\ref{dos}), and is
clearly a finite size effect. The lower energy scale is associated
with the additional onsite screening, due to the presence of other
local spins in the lattice. The intersite \SScO{} correlations are
still approximately governed by a single energy scale, $J{\rm_K}^2/t$,
just as for PBC. Indeed the behavior of the normalized intersite
correlation functions for OBC and PBC are quite similar,
despite of the fact that the magnitude differs by a factor of two.

\section{Magnetic susceptibility and total local magnetic moment}

\begin{figure}
\begin{center}
\includegraphics[width=0.9\linewidth,clip=]{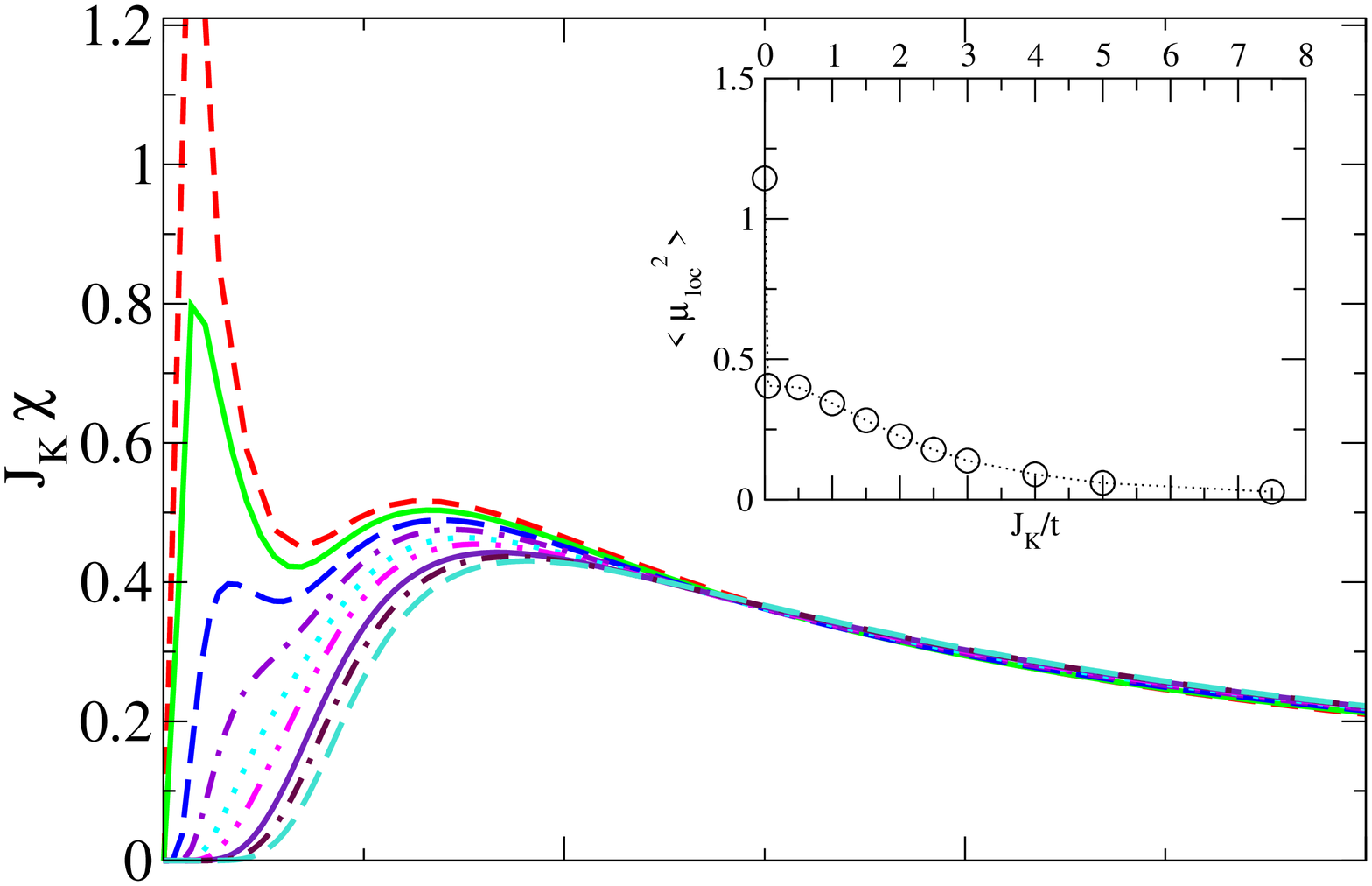}
\includegraphics[width=0.9\linewidth,clip=]{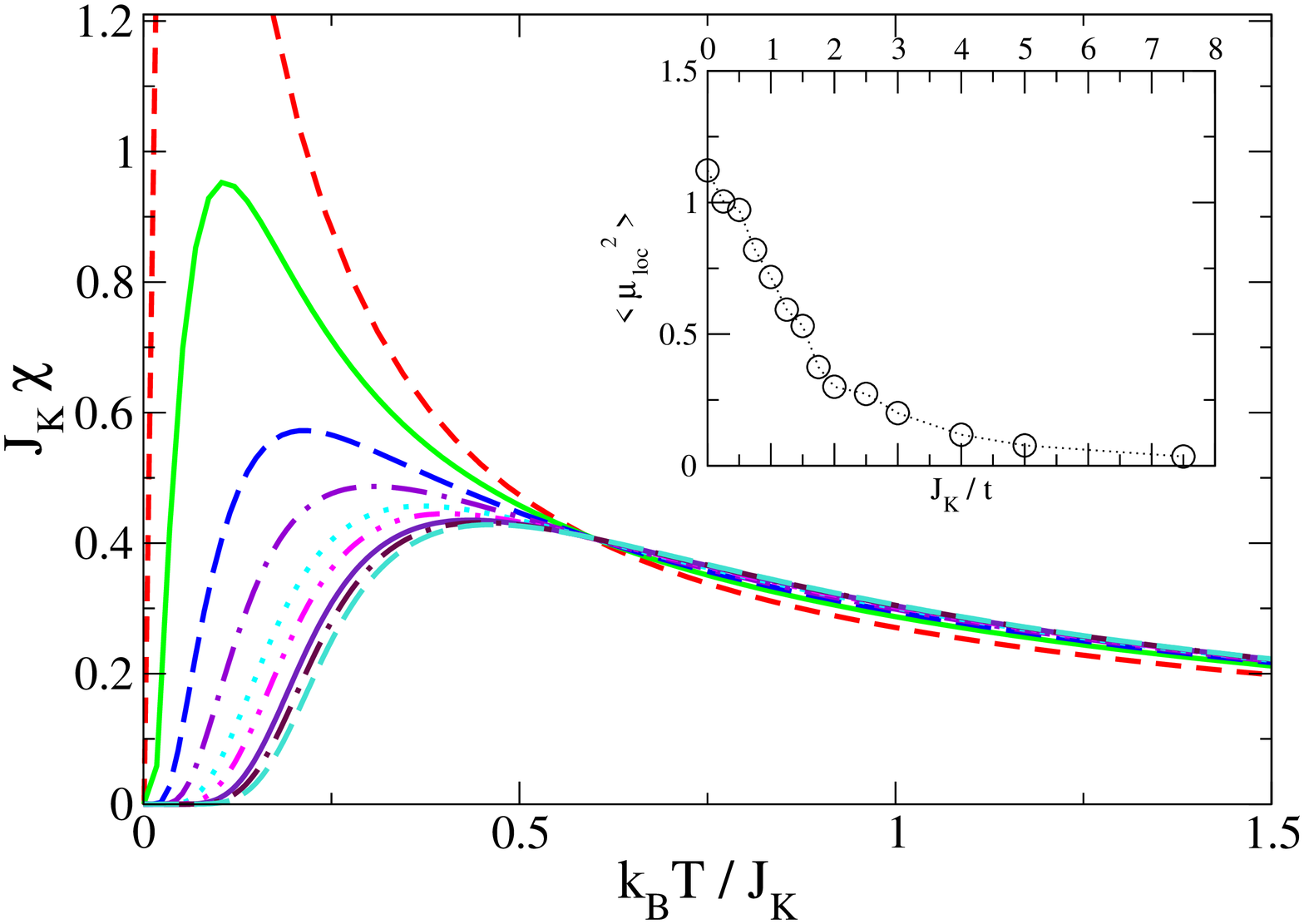}
\caption{\label{chi} (Color online) The scaled magnetic susceptibility, as indicated
on the axes. Above: for PBC. Below: for OBC. In the insets the total local
moments in the ground state (at $T=0$~K) as function of \JKt{} are shown,
$\langle\bm\mu_{\rm loc}^2\rangle=\langle(\bm\tau_i+{\mathbf S}_i)^2\rangle$,
($i=1,3$ for the PBC and OBC, respectively).}
\end{center}
\end{figure}

The spin correlations are also directly reflected in
magnetic susceptibility, shown in Fig.~\ref{chi}, for PBC and OBC.
It is multiplied by \JK{} and plotted versus temperature scaled with
\JK, as was done in Ref.~\onlinecite{Capponi:2001}. The two peaks for PBC
are clearly seen for small \JKt, as in the specific heat. However, in the magnetic susceptibility even the lower peak for \JKt$=0.5$ is clearly seen. This is because the lower peaks in the magnetic susceptibility are much more pronounced and at slightly higher temperatures than the corresponding peaks in the specific heat. The position of the lower peak scales well with $J\rm_K^2${} (not shown in the figure).  The lower peak is not distinguishable any more for $J_{\rm K}/t\gtrsim2$. The position of the higher 
peak scales with \JK. For OBC only one large peak is seen. For $J_{\rm K}/t\lesssim2${} it scales approximately with $J\rm_K^2$. In the strong coupling limit, $J_{\rm K}/t\gtrsim2$, the
peak scales linearly with \JK, just as for PBC. The results for OBC are similar to the results for the 10-site cluster with PBC in Ref.~\onlinecite{Haule:2000}. These results are also similar to the results of the QMC calculations for larger clusters in Ref.~\onlinecite{Capponi:2001}. However, in Ref.~\onlinecite{Capponi:2001},  the magnetic susceptibility peak scales with \JK$^2$, while in the onsite correlations the characteristic temperature is proportional to \JK, suggesting the existence of two low energy scales.

In the insets of Fig.~\ref{chi} the ground state values of the total local moments are
shown, defined as:
\begin{equation}\label{mu}
\langle\bm\mu_{\rm loc}^2\rangle=\langle({\bm\tau}_i+\mathbf
S_i)^2\rangle=\langle\bm\tau_i^2\rangle+\langle{\mathbf S}_i^2\rangle +
2\langle\bm\tau_i\cdot\mathbf{S}_i\rangle.
\end{equation}
The interesting quantity here is $\langle\bm\tau_i^2\rangle$ which is a measure of itinerant spin localization. In the strong coupling limit $\langle\bm\tau_i^2\rangle=3/4$, i.e., the itinerant spins become localized as the onsite singlets are formed and the hopping of conduction electrons is frozen. When the conduction electrons are completely delocalized, i.e., for \JK$=0$, $\langle\bm\tau_i^2\rangle=3/8$, because the probability of finding one electron with spin up or down on one particular site is $1/2$. Accordingly, $\langle\bm\mu_{\rm loc}^2\rangle=\langle\bm\mu_{\rm loc}^2\rangle_0=9/8$ for $J_{\rm K}=0$ while for $J_{\rm K}\rightarrow\infty$ it vanishes, as the onsite singlets are formed  and the local moments completely screened. There is a jump for \JKt$>0$ for PBC. It has the same origin as the jump for the onsite correlations \zSc{}, i.e., the highly degenerate unperturbed conduction electron ground state. A jump is also present in $\langle\bm\tau_i^2\rangle$. For  OBC we see a continuous decrease of $\langle\bm\mu_{\rm loc}^2\rangle$.

The $\langle\bm\mu_{\rm loc}^2\rangle$ is a good quantity to define a critical value of \JKt{} at which the crossover from predominantly magnetic to predominantly nonmagnetic ground state occurs. The other possibility would be to compare directly the intersite and onsite correlations, but the intersite correlations depend strongly on the cluster size. Therefore we define a critical (\JKt{})$_{\rm C}$ as the value for which $\langle\rm\mu_{loc}^2\rangle=\langle\rm\mu_{loc}^2\rangle_0/2$. Then
for OBC, (\JKt{})$_{\rm C}\approx 1.4$ (cf. Fig.~\ref{chi}), in
good agreement with the values 
obtained from the Quantum Monte Carlo~\cite{Capponi:2001} and the
mean-field analysis.~\cite{Zhang:2000} For PBC it would be $0$, but
for this small cluster size, OBC are much more representative for
the thermodynamic limit of the KLM.

\section{Phase diagram and conclusions}

\begin{figure}
\begin{center}
\includegraphics[width=0.9\linewidth,clip=]{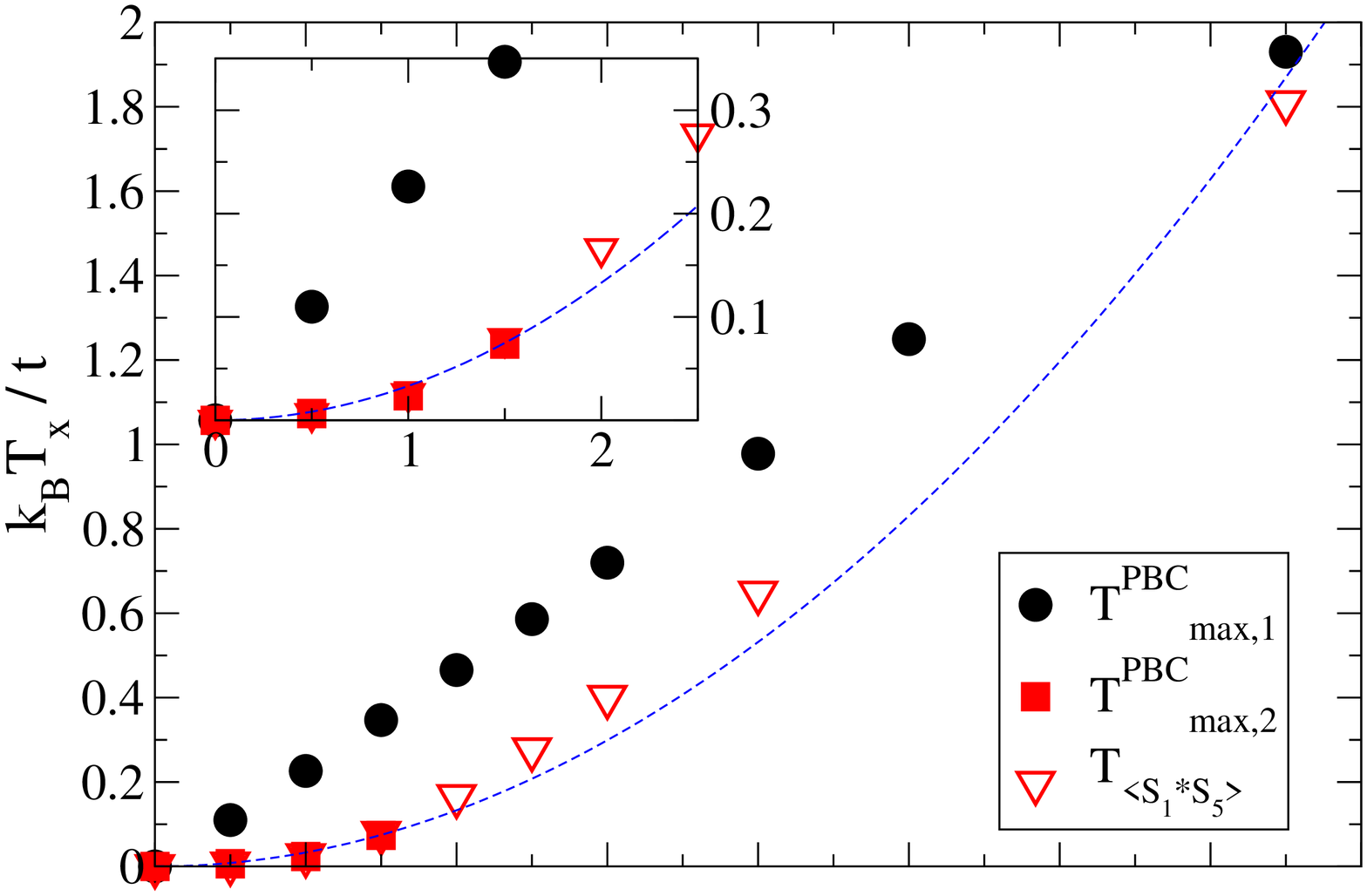}
\includegraphics[width=0.9\linewidth,clip=]{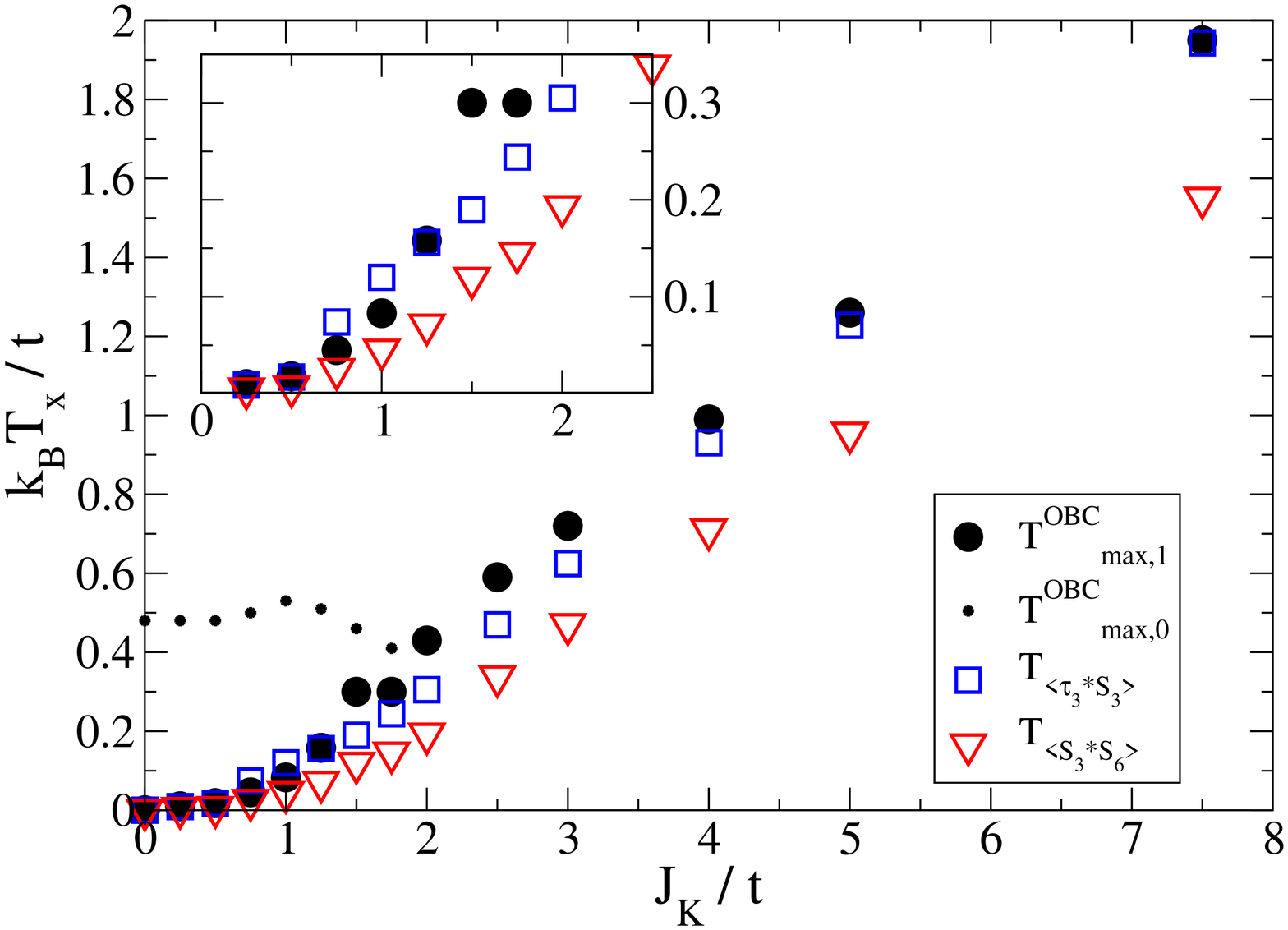}
\caption{\label{phdia} (Color online) The phase diagrams for the 8-site 2D clusters. In the insets the weak coupling region is magnified.
Above: for PBC. The inflection points are approximately determined as the temperatures where the normalized \SSc{} has the value of -0.75. The dashed line is the fitting curve assuming quadratic dependence on \JK. The inflection points of the onsite correlations agree with the positions of the upper peak in the specific heat and are not shown in the Figure. Below: for OBC. The inflection points are approximately determined as the temperatures where the
normalized correlation functions achieve the values of -0.8 and -0.75 for \zScO{}
and \SScO{}, respectively.}
\end{center}
\end{figure}

We may summarize the results in form of phase diagrams for PBC and OBC, shown in Fig.~\ref{phdia}.  In the phase diagrams the positions of specific heat maxima are plotted as function of coupling strength, \JKt. Also the inflection points of the correlation functions are plotted, and they mostly follow the positions of the corresponding specific heat maxima.

In the weak coupling regime, for PBC, two characteristic temperatures, $T\rm_{max,1}^{PBC}$ and $T\rm_{max,2}^{PBC}$, are found. The $T\rm_{max,2}^{PBC}$ cannot be determined for \JKt$\gtrsim2$, but we make the continuation with the \SSc{} inflection point temperature, $T\rm_{\langle\mathbf S_1\cdot\mathbf S_5\rangle}$, which can be defined for all coupling strengths. Therefore we conclude that the corresponding anomaly in the specific heat is still there, but is too small to be identified. The inflection points of the correlation functions reveal that the characteristic temperatures $T\rm_{max,1}^{PBC}$ and $T\rm_{max,2}^{PBC}$ correspond to the energy scales of onsite correlations and intersite correlations, respectively. The former is linear in \JK{} for all \JKt. The latter is proportional to $J{\rm_K}^2/t$, and may be associated with the RKKY interaction. 

For OBC, there is
only one characteristic temperature, $T\rm_{max,1}^{OBC}$, for all \JKt. In the weak coupling regime, for \JKt$\lesssim2$, below $T\rm_{max,1}^{OBC}$ predominantly intersite correlations are formed. However, $T\rm_{max,1}^{OBC}$ is also the characteristic temperature of onsite correlations. This is seen from the positions of inflection points for the corresponding correlations, also shown in the lower part of Fig.~\ref{phdia}. Because the intersite correlations dominate and the characteristic temperature is approximately quadratic in \JK{}, the corresponding energy scale, in the weak coupling regime, may be associated with the RKKY interaction. 

It is important to note that $T{\rm_{max,1}^{OBC}}\approx T\rm_{max,2}^{PBC}$, i.e., the RKKY interaction is not very sensitive to the distribution of non-interacting conduction electron states. This is plausible with the known expression for the RKKY characteristic temperature, $T_{\rm
RKKY}\propto J_{\rm K}^2/W$ (e.g.~\cite{Sun:2005}), where $W$ is the band
width. It is also important that the RKKY energy scale is the energy scale of the onsite correlations in the weak coupling regime for the OBC.  We may generally conclude that the RKKY energy scale is the lowest energy scale of the KLM; if the local spins are not screened above $T\rm_{RKKY}$,
then it is the only energy scale of the system. 

In contrast, the onsite correlations, i.e., the onsite screening of local spins, is very
sensitive to the number of states within \JK{} around the "Fermi
level", as seen from the difference between $T\rm_{max,1}^{PBC}$ and $T\rm_{max,1}^{OBC}$  for PBC and OBC, respectively. This is plausible with the expression for the single impurity Kondo 
temperature, which depends exponentially on the DOS at the Fermi
level, $\rho(\epsilon_{\rm F})$; $T_{\rm K}=\epsilon_{\rm
F}\exp{\left[-1/(J\rho(\epsilon_{\rm F}))\right]}$.

In the strong coupling regime, only one characteristic temperature is found for both PBC and OBC. It corresponds to the energy of the onsite singlet formation, which is linear in \JK. Boundary conditions are unimportant, because the local physics determines the properties of the system.

We may argue on the relevance of these results for the system in the
thermodynamic limit. Recently a QCP was found
\cite{Schroder:2000,Custers:2003,Paschen:2004}  whose properties
deviate from the expected behavior
(e.g.~\cite{Steglich:1996,Mathur:1998}). The properties of the system
at the QCP are determined by the competition between the onsite
screening and the RKKY interaction. Here we have shown that the
competition between the onsite screening and the RKKY interaction
strongly depends on the particular form of the distribution of
non-interecting conduction electron levels. We may anticipate that in
the thermodynamic limit this means a strong dependence on the
particular form of non-interacting conduction electron DOS. A
strongly peaked DOS near the Fermi level favors onsite screening,
while the RKKY interaction is not very sensitive to the particular
shape of the DOS. These results indicate that the shape of the DOS is
very important for the properties of the heavy fermion systems in the
vicinity of the QCP.

Finally, the qualitative behavior of the
correlation functions for the KLM with OBC is similar to the KLM
without charge degrees of freedom,~\cite{Zerec:2005b} which is a
generalized Kondo necklace model in 2D from Doniach's original
work.~\cite{Doniach:1977} This shows that this simplified model describes well the spin degrees of freedom in the KLM at half-filling.

\begin{acknowledgments}

We wish to acknowledge useful discussions with P. Fulde, S. Burdin, N. Perkins, and R. Ramazashvili.

\end{acknowledgments}


\end{document}